\documentclass[aps,prl,preprint,groupedaddress,showpacs,showkeys]{revtex4-1}
\usepackage{amsmath}
\usepackage{gensymb}
\usepackage{graphicx}
\usepackage{epstopdf} %converting to PDF
\usepackage{xr}
\usepackage{nth}
\usepackage{lipsum}
\usepackage{physics}
\usepackage{longtable}

\newcommand{\ignore}[1]{}
\begin{document}

%\preprint{}

\title{Finding dominant reaction pathways via global optimization of action}


\author{Juyong Lee}
%\email[]{juyong.lee@nih.gov}
%\thanks{}
\affiliation{Laboratory of Computational Biology, National Heart, Lung, and Blood Institute (NHLBI), National Institutes of Health (NIH), Bethesda, Maryland 20892, U.S.A.}

\author{In-Ho Lee}
\affiliation{Korea Research Institute of Standards and Science, Daejeon 34113, Republic of Korea}

\author{InSuk Joung}
\affiliation{School of Computational Sciences, Korea Institute of Advanced Study, Seoul 02455, Korea}

\author{Jooyoung Lee}
\email[Co-corresponding author: ]{jlee@kias.re.kr}
\affiliation{School of Computational Sciences, Korea Institute of Advanced Study, Seoul 02455, Korea}

\author{Bernard R. Brooks}
\email[Co-corresponding author: ]{brb@nih.gov}
\affiliation{Laboratory of Computational Biology, National Heart, Lung, and Blood Institute (NHLBI), National Institutes of Health (NIH), Bethesda, Maryland 20892, USA}


\date{\today}

\begin{abstract}

We present a new computational approach to sample multiple reaction pathways 
with fixed initial and final states through global optimization 
of the Onsager-Machlup action using the conformational space annealing method.
This approach successfully samples not only the most dominant pathway 
but also many other possible paths without initial guesses on reaction pathways.
Pathway space is efficiently sampled by crossover operations of a set of paths and preserving the diversity of sampled pathways.
The sampling ability of the approach is assessed by finding pathways 
for the conformational changes of alanine dipeptide and hexane.
The benchmarks demonstrate that the rank order and the transition time distribution of multiple pathways 
identified by the new approach are in good agreement with those of long molecular dynamics simulations.
We also show that the folding pathway of the mini-protein FSD-1 identified by the new approach 
is consistent with previous molecular dynamics simulations and experiments.
\end{abstract}

%\keywords{}

\maketitle

%%\section{\label{Introduction}Introduction}
%%\subsection{}
%%\subsubsection{}

Finding multiple reaction pathways between two end states 
remains a challenging problem in computational biophysics~\cite{Elber2016}. %%despite of recent progresses in 
For this purpose, performing a long-time molecular dynamics (MD) simulation is a commonly used approach. 
Despite recent progress in the methodologies of MD, this approach still suffers from a timescale problem. 
Many biological reactions such as protein folding and conformational transitions occur in 
the microsecond or millisecond ranges, which are hard to be performed with conventional computers. 
In addition, MD simulations starting from one end state are not guaranteed to reach the other end state of interest.
Thus developing an efficient computational method to find multiple possible reaction pathways connecting two end states is necessary.
There are currently no methods that can efficiently produce the multiple dominant pathways 
connecting two well-defined end point states in a complex system.
The objective of this work is to present such a method.
Other approaches using a conformational driving force do not sample alternatives.
Methods that are robust, such as transition path sampling~\cite{Dellago1998,Bolhuis2002}, are very expensive to use for complex systems with and multiple steps and barriers.

Various chain-of-state methods have been suggested based on the assumption 
that a dominant transition pathway between two states follows the minimum energy pathway~\cite{Czerminski1990,Henkelman2000,E2002}. %starting from an initial guess.
The limitations of these methods are that they do not consider the dynamics of a system 
and find only the nearest local minimum solution from a given initial pathway~\cite{Elber2016}.
Alternative methods based on the principle of least action have been suggested~\cite{Gillilan1992,Olender1996,Elber1999,Elber2000,Passerone2001,Passerone2003,Lee2003a}.
Passerone and Parrinello suggested the action-derived molecular dynamics (ADMD) method 
based on the combination of classical action and a penalty term that conserves the total energy of a system~\cite{Passerone2001,Passerone2003}.
To enhance the convergence of ADMD calculations, Lee et al. introduced 
a kinetic energy penalty term based on the equipartition theorem~\cite{Lee2003a,Lee2005,Lee2012c,Lee2013c}.
Although ADMD yields physically relevant pathways, it has two practical limitations~\cite{Crehuet2003,Lee2003a}: 
i) it strongly depends on the initial guess of a pathway; 
and ii) it cannot identify the most dominant pathway when there are multiple pathways 
because the classical least action principle is an extremum principle.

For diffusive processes, the second problem can be avoided 
by using the Onsager-Machlup (OM) action $S_\text{OM}$~\cite{Onsager1953,Machlup1953,Olender1996,Eastman2001,Zuckerman2001,Fujisaki2010,Fujisaki2013}.
Onsager and Machlup showed that the relative probability to observe a pathway 
with an OM action of $S$ is proportional to $e^{-S/k_\text{B}T}$ where $k_\text{B}$ is the Boltzmann constant and $T$ is a temperature. %%is a diffusion constant. 
Thus the most dominant pathway corresponds to the one that minimizes $S_\text{OM}$ 
and the same result can be obtained by solving the Fokker-Planck equation~\cite{Faccioli2006,Sega2007,Beccara2012}. 
This property recasts the problem of finding dominant pathways into a global optimization problem.
However, finding the global minimum of $S_\text{OM}$ is a numerically challenging task 
because the minimization of $S_\text{OM}$ requires the second derivatives of a potential function, 
which are computationally expensive, at best, and wholly unavailable for may quantum mechanical energy surfaces.

In this work, we propose an efficient computational method that finds not only the most dominant pathway 
but also multiple suboptimal pathways without second derivative calculations.
For global optimization of $S_\text{OM}$, we used an efficient global optimization method called conformational space annealing (CSA) 
based on a combination of genetic algorithm, simulated annealing, and Monte Carlo minimization~\cite{Lee1997,Lee1999b}.
%The CSA method has been demonstrated to outperform other algorithms, e.g. simulated annealing~\cite{Scott1983}, 
The CSA method has been demonstrated to be extremely efficient in solving various global optimization problems including 
finding low energy conformations of Lennard-Jones clusters~\cite{Lee2003},
protein structure prediction~\cite{Lee1999b,Lee2003,Steinbach2004,Joo2007,Lee2008,Joo2009,Lee2011c}, 
multiple sequence alignment~\cite{Joo2008}, 
and community detection in networks~\cite{Lee2012b,Lee2013a,Lee2013b}.
We extend the CSA approach to examining pathways, preserving all features 
that make it robust and efficient, by applying it to sets of entire pathways represented as a chain-of-states.
From benchmark simulations using alanine dipeptide, we observed that our method finds multiple transition pathways, 
which are consistent with long-time Langevin dynamics (LD) simulations.
In addition, the rank order statistics and transition time distributions of the multiple transition pathways 
are in good agreement with those of the LD results.
We will call the method Action-CSA.

Here, we briefly review the theoretical background behind Action-CSA.
If a system with $N$ atoms with a potential energy $V$ follows the overdamped Langevin dynamics, 
\begin{equation}
  \gamma \mathbf{m}\mathbf{\dot{x}} = -\frac{\partial V}{\partial \mathbf{x}} + \mathbf{R},
  \label{eq:overdamped_langevin}
\end{equation}
where $\mathbf{m}$ is a diagonal mass matrix, $\mathbf{x}$ is a $3N$ dimensional coordinate vector, 
$\gamma$ is collision frequency,
and $\mathbf{R}$ is a Gaussian random force, 
the relative probability of finding a final state $\mathbf{x}_\textup{f}$ at a time $t$ 
from an initial state $\mathbf{x}_\textup{i}$ via diffusive trajectories $\mathbf{x}(t)$ 
is determined by using the path integral approach and OM action $S_\text{OM}[\mathbf{x}(t)]$~\cite{Onsager1953,Machlup1953}:
\begin{equation}
  %%P(\mathbf{x}_f|\mathbf{x}_i;t) = \int_{0}^{t} \mathcal{D}\mathbf{x}(t) \text{exp}\left(-\frac{S_\text{OM}[\mathbf{x}(t)]}{k_\text{B} T} \right).
  P(\mathbf{x}_\textup{f}|\mathbf{x}_\textup{i};t) = \int_{0}^{t} \mathcal{D}\mathbf{x}(t) \; \text{exp} \left(-\frac{S_\text{OM}[\mathbf{x}(t)]}{k_\text{B} T} \right),
  \label{eq:traj_prob}
\end{equation}
where $\mathcal{D}\mathbf{x}(t)$ indicates that the integration runs over all possible pathways $\mathbf{x}(t)$.
%%Due to this relationship, the pathway corresponding to the global minimum of $S_\text{OM}$ is the most dominant 
This relationship suggests that if the $S_\text{OM}$ values of all physically 
accessible pathways are obtained, one can determine the relative populations of multiple pathways.
Thus, $S_\text{OM}$ is a proper target objective function of global optimization.
The generalized OM action of a pathway $\mathbf{x}(t)$ is defined~\cite{Onsager1953,Machlup1953,Hunt1981,Adib2008}:
\begin{widetext}
\begin{equation}
  S_\text{OM}[\mathbf{x}(t)] = \frac{\Delta V}{2} + \frac{1}{4 \gamma} \int_{0}^{t} d\tau \left\{ \left[ \gamma \mathbf{m} \mathbf{\dot{x}}(\tau) \right]^2 + \norm{\nabla V[\mathbf{x}(\tau)]}^2 - 2 k_\text{B}T \nabla^2 V[\mathbf{x}(\tau)] \right\},
    \label{eq:om_ac1}
\end{equation}
\end{widetext}
where $\Delta V = V(\mathbf{x}_\textup{f})-V(\mathbf{x}_\textup{i})$.
The last term is related to trajectory entropy connected with fluctuations~\cite{Adib2008,Haas2014a,Haas2014b} 
and was not presented in the orignal work by Onsager and Machlup because the harmonic potential was considered~\cite{Onsager1953,Machlup1953}.
Note that the minimization of $S_\text{OM}$ using analytic local minimization algorithms requires third derivatives.
This makes the direct global optimization of $S_\text{OM}$ hard to be applied to detect 
transition pathways of biomolecules with all-atom force fields due to the complexity of implementation and high computational cost.
For numerical calculations based on a chain-of-state representation, the OM action should be discretized. 
The method uses the second-order discretization of the symmetric OM formula, which 
uses only gradients for $S_\text{OM}$ calculations~\cite{Miller2007}:
%\lipsum[1]
\begin{widetext}
\begin{equation}
  \begin{split}
    S_\text{OM}[\mathbf{x}(t)] = \frac{\Delta V}{2} + \sum_{i=0}^{P-1} \frac{\Delta t}{4\gamma} \left\{ \left[ \frac{ \gamma \mathbf{m} (\mathbf{x}_{i+1}-\mathbf{x}_{i})}{\Delta t} \right]^2 + \frac{ \norm{ \nabla V(\mathbf{x}_i) }^2 + \norm{ \nabla V(\mathbf{x}_{i+1}) }^2}{2} \right. \\
      \left. - \frac{\gamma \mathbf{m} (\mathbf{x}_{i+1}-\mathbf{x}_{i})}{\Delta t} [ \nabla V(\mathbf{x}_{i+1}) - \nabla V(\mathbf{x}_{i}) ] \right\},
      %%\left. - \left( \frac{\gamma \mathbf{m} (\mathbf{x}_{i+1}-\mathbf{x}_{i})}{\Delta t} \right) (\nabla V(\mathbf{x}_{i+1}) - \nabla V(\mathbf{x}_{i})) \right].
  \end{split}
    %% S_\text{OM}[\mathbf{x}(t)] = \frac{\Delta V}{2} + \sum_{i=0}^{P-1} \frac{\Delta t}{4\gamma} \left[ \left( \frac{ \gamma \mathbf{m} (\mathbf{x}_{i+1}-\mathbf{x}_{i})}{\Delta t} \right)^2 + \frac{(\nabla V(\mathbf{x}_i))^2 + (\nabla V(\mathbf{x}_{i+1}))^2}{2} - \frac{\gamma \mathbf{m} (\mathbf{x}_{i+1}-\mathbf{x}_{i})}{\Delta t} (\nabla V(\mathbf{x}_{i+1}) - \nabla V(\mathbf{x}_{i})) \right],
  \label{eq:disc_om}
\end{equation}
\end{widetext}
%\lipsum[1]
where $P+1$ is the number of replicas, $\Delta t$ is a time step between successive replicas, and $t = P\Delta t$ is the total transition time.
This formula is superior to the direct implementation of Eq. (\ref{eq:om_ac1}) 
since it requires only the first derivatives of $V$ to evaluate $S_\text{OM}$.

%%If a system of interest has $K$ degrees of freedom, a configuration is represented as a sequence of $K$ real values. 
%%finding dominant pathways becomes a global optimization problem with $3NM$ degrees of freedom.
Here, we describe the application of CSA to optimize $S_\text{OM}$. %%the overall procedure of CSA.
In general, a pathway is represented as a chain of $P-1$ replicas with $N$ atoms for each replica leading to $3N(P-1)$ total degrees of freedom. 
Each replica is represented by a sequence of $3N-6$ internal dihedral angles and 6 net translational/rotational degrees of freedom.
An Action-CSA calculation starts with a set of random pathways on a pathway space. 
Subsequently, the actions of the random pathways should be locally optimized.
We call this set of pathways a bank, and update conformations in the bank during the Action-CSA calculation.
As stated previously, direct minimization of $S_\text{OM}$ using analytic gradients is computationally challenging.

For a computationally feasible local action optimization, we optimized a pathway using a modified action from ADMD instead of using $S_\text{OM}$.
The discretized classical action is defined: %%of the \emph{j}th step is defined as 
\begin{widetext}
\begin{equation}
  S_\text{classical}[\mathbf{x}(t)] = \sum_{i=0}^{P-1} L_i(\mathbf{x}_i) \Delta t = \sum_{i=0}^{P-1} \left[ \frac{\mathbf{m} (\mathbf{x}_{i} - \mathbf{x}_{i+1})^2}{2\Delta t^2} - V(\mathbf{x}_{i}) \right] \Delta t.
  \label{eq:classical_action}
\end{equation}
\end{widetext}
%%The principle of classical action is finding stationary points of $S_{classical}$. 
Physically accessible pathways correspond to the stationary points of $S_\text{classical}$.
Finding such pathways is a computationally difficult task because $S_\text{classical}$ is not bounded;
$S_\text{classical}$ can be minimized or maximized, and the stationary points of $S_\text{classical}$ can be minima, maxima or saddle points.
Another practical problem is that the total energies of pathways satisfying the stationary condition $\delta S_\text{classical}=0$ may not be conserved~\cite{Passerone2001}.
To find pathways that satisfy the principle of least action and conserve total energies, a modified action with a penalty term restraining total energy was suggested~\cite{Passerone2001}:
%%To obtain a physically allowed pathway from a trial pathway, the ADMD approach~\cite{Passerone2001} is used. %%Lee et al. suggested the kinetically controlled action~\cite{Lee2003a}:
%%+ \mu_{K} \sum_{i=1}^{N}\left( \langle K_i \rangle - \frac{3k_\text{B}T}{2}\right)^2
\begin{widetext}
\begin{equation}
  \begin{split}
    \Theta(\mathbf{x}_{i} ; E) & = \mu_{A} S_\text{classical} + \mu_{E} \sum_{i=0}^{P-1}(E_i - E)^2 \\ 
    & = \mu_{A} \sum_{i=0}^{P-1} \left[ \frac{\mathbf{m} (\mathbf{x}_{i} - \mathbf{x}_{i+1})^2}{2\Delta t^2} - V(\mathbf{x}_{i}) \right] \Delta t
    + \mu_{E} \sum_{i=0}^{P-1} \left\{ \left[ \frac{\mathbf{m} (\mathbf{x}_{i} - \mathbf{x}_{i+1})^2}{2\Delta t^2} + V(\mathbf{x}_{i}) \right] - E \right\}^2 ,
  \end{split}
    \label{eq:mod_ac2}
\end{equation}
\end{widetext}
where $E$ is a targeted total energy of a system, 
$\mu_A$ and $\mu_E$ are the weighting parameters of the classical action, 
and the restraint term for energy conservation.
The minimization of $\Theta[\mathbf{x}(t);E]$ requires only the first derivatives of $V$.
%%where $\langle K_i \rangle$ is the average kinetic energy of $i$th atom along the pathway and $\mu_E$ is the weighting parameter of the total energy conservation restraint term.

We call the set of locally optimized initial random pathways using $\Theta[\mathbf{x}(t);E]$ the \emph{first bank}.
The first bank remains the same throughout the optimization and is used 
as the reservoir of partially optimized pathways to enhance the diversity of pathway search.
A copy of the first bank is generated and called a \emph{bank}.
The conformations in the bank are updated during a calculation 
while the size of the bank is kept constant.
By using the pathways included in the first bank and the bank, 
new trial pathways are generated by performing crossover and random perturbation operations.
For a crossover operation, two pathways, a seed pathway from the bank and a random pathway either from the bank or the first bank, are selected 
and random parts of two selected configurations are swapped.
%%For a crossover operation, a certain number of seed pathways from the bank and random pathways from the bank or the first bank are selected.
%%In addition to crossover operations, random perturbations are performed.
For a random perturbation, a certain number of degrees of freedom of a seed pathway, up to 5\% of total degrees of freedom, were randomly changed.
%%The number of perturbations was selected randomly with the maximum of 
The generated trial pathways are locally optimized using $\Theta[\mathbf{x}(t);E]$ 
to remove any possible artifacts generated by the crossover and the random perturbation operations.
However, after local minimizations, the bank was updated by comparing the $S_\text{OM}$ values 
of the existing pathways and the new ones instead of $\Theta[\mathbf{x}(t);E]$.

A key feature of CSA is a sophisticated bank-update procedure 
that prevents a search being trapped in local minima during the optimization 
and keeps the diversity of the bank.
%%the bank was updated while keeping the diversity of the bank.
For a newly obtained configuration $\alpha$, the distances between $\alpha$ and the existing ones in the bank are calculated.
If the distance $D$ between $\alpha$ and its closest neighbor is less than a cutoff distance $D_\text{cut}$, 
only the better configuration in terms of the objective function is selected.
If $D>D_\text{cut}$, $\alpha$ is considered a new configuration and it replaces the worst configuration in the bank.
At initial stages of a calculation, $D_\text{cut}$ is kept large for wider sampling.
As the calculation proceeds, it gradually decreases for a refined search near the global minimum. 
The bank keeps updating until no pathway with a lower $S_\text{OM}$ is found.
In this work, a distance between two pathways was measured by the Fr\'{e}chet distance~\cite{alt1995}.
More details on a general CSA procedure are described elsewhere~\cite{Lee1997,Lee1999b,Lee2003,Joo2008,Joo2009,Lee2011c}.

%%% Results and Discussion %%%
%% Method part %%
To verify that Action-CSA successfully finds multiple transition pathways and allows one to determine 
the rank order of the pathways based on their optimized $S_\text{OM}$ values, 
we applied our method to investigate the conformational transition of alanine dipeptide from $C7_\text{eq}$ to $C7_\text{ax}$ in the vacuum.
Here, we used the polar hydrogen representation in the PARAM19 force field~\cite{Neria1996} and dielectric constant was set to 1.0~\cite{Loncharich1992}.
We performed Action-CSA simulations with different transition times, $t$ in Eqs. (\ref{eq:disc_om}) and (\ref{eq:mod_ac2}), 
ranging from 0.2 ps to 2.0 ps with an interval of 0.1 ps.
%%The numbers of replicas were adjusted according to transition time to keep the time step between replicas $\Delta t$ = 5 fs.
The numbers of replicas were adjusted with $t$ to keep the time step between replicas $\Delta t$ = 5 fs.
All simulations were performed at temperature $T$ = 350 K with a collision frequency $\gamma$ = 1.0 ps$^{-1}$.
The reference total energy $E$ in Eq. (\ref{eq:classical_action}) was obtained 
by adding the initial potential energy $V(\mathbf{x}_i) = -43.3$ kcal/mol and a kinetic energy of 12.5 kcal/mol 
estimated by $3Nk_\text{B}T/2$ with the number of atoms $N$ = 12.
The weighting parameters $\mu_{A}$ and $\mu_{E}$ in Eq. (\ref{eq:classical_action}) were set to $-1.0$ and $1.0$, respectively.
For comparison purposes, we performed 500 $\mu$s LD simulations of alanine dipeptide under the same condition 
and counted the number of the $C7_\text{eq}$ $\to$ $C7_\text{ax}$
 transitions.

%%%%%%%%%%%%%%%%%%%%%%%%%%%%%%%%%%%%%%%%%%%%%%%%%%%%%%%
%%%%%%%%%%%%%%%%% Result part %%%%%%%%%%%%%%%%%%%%%%%%%
%%%%%%%%%%%%%%%%%%%%%%%%%%%%%%%%%%%%%%%%%%%%%%%%%%%%%%%

\begin{figure}[!htbp]
  \includegraphics[width=0.70\columnwidth]{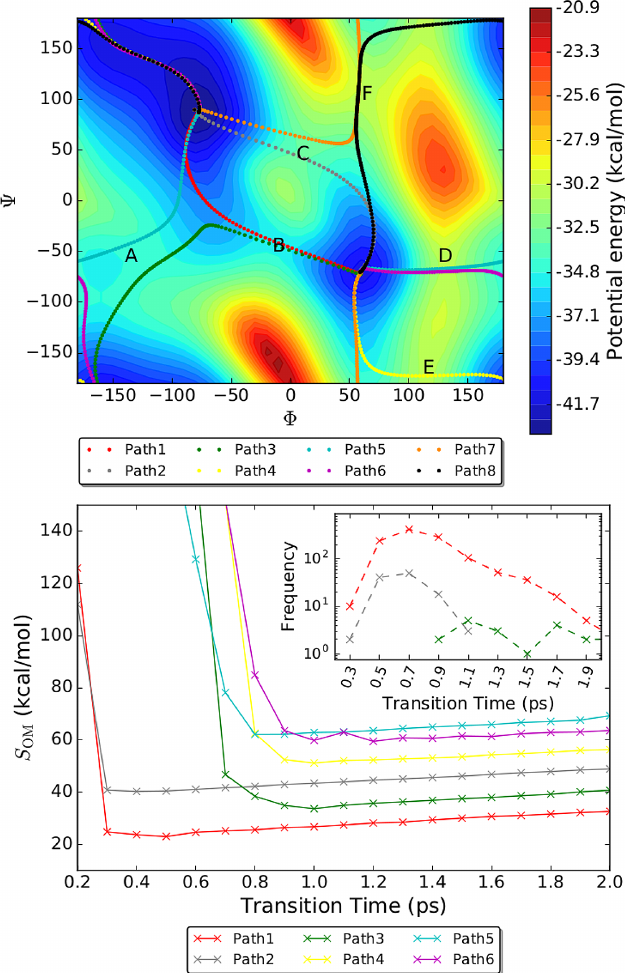}
  \caption{Upper panel: eight different pathways for the $C7_\text{eq}$ $\to$ $C7_\text{ax}$ transition 
    and the potential energy surface for the $\Phi$ and $\Psi$ angles with the PARAM19 force field (in units of kcal/mol).
    Potential energy barriers are labeled in order of their heights (from A to F).
    Lower panel: the $S_\text{OM}$ values of 6 pathways for the $C7_\text{eq}$ $\to$ $C7_\text{ax}$ 
    transitions of alanine dipeptide along different transition times.
  }
  \label{fig:alanine_dipeptide}
\end{figure}

Now we will show that the Action-CSA identifies not only the most dominant pathway but also multiple possible pathways. %%but also information on transition times.
We identified 8 different pathways for the $C7_\text{eq}$ $\to$ $C7_\text{ax}$ transition 
by clustering all pathways sampled from the Action-CSA simulations (Figure~\ref{fig:alanine_dipeptide}A).
From the $S_\text{OM}$ values obtained with different transition times (Figure~\ref{fig:alanine_dipeptide}B), 
it is clear that the pathway that passes barrier B has the lowest $S_\text{OM}$ values along all transition times tested, 
which indicates that it is the most dominant pathway regardless of transition time.
This is consistent with the LD simulation results. 
To compare the Action-CSA result with LD, we performed 5000 independent 100 ns LD simulations amounting to 500 $\mu$s trajectories (Table \ref{tab:ld_result}). %\ref{Sptab:ld_result}).
From the simulations, we observed 1350 transitions from $C7_\text{eq}$ to $C7_\text{ax}$ 
and categorized them by finding the nearest neighbor from the 8 pathways obtained with CSA.
Among them, the pathway that crosses barrier B was identified as the most dominant one with all transition times
This demonstrates that Action-CSA correctly identified the minimum OM action pathway and that it matches the most dominant pathway observed in LD simulations.

\begin{center}
\begin{table}[htbp!]
  \caption[table1_caption]{The frequences of pathways observed from 500 $\mu$s Langevin dynamics simulations.}
  \label{tab:ld_result}
  \begin{tabular}{c|c} %add [H] placement to break table across pages
      \hline \hline
      Path ID  & Frequency       \\ \hline
      Path1 & 1183 \\
      Path2 & 116  \\
      Path3 & 25   \\
      Path4 & 7    \\
      Path5 & 4    \\
      Path6 & 4    \\
      Path7 & 10   \\
      Path8 & 1    \\
      %% Path1    & 1036 $\pm$ 10.6 \\
      %% Path2    &   89 $\pm$ 8.9  \\
      %% Path3    &   25 $\pm$ 4.8  \\
      %% Path4    &    5 $\pm$ 2.1  \\
      %% Path5    &    7 $\pm$ 2.7  \\
      %% Path6    &    8 $\pm$ 3.0  \\
      \hline
  \end{tabular}
\end{table}
\end{center}

%%However, identifying the order of less dominant pathways from the $S_\text{OM}$ values is not straightforward because the order of $S_\text{OM}$ changes with $t$.
%%A comparison of the LD results and the Action-CSA results also shows that the CSA results
In addition, it is also identified that the Action-CSA simulations 
can provide information on the transition times of various pathways.
Until $t < 0.8$ ps, the pathway that crosses barrier C (Path2) has 
the second lowest $S_\text{OM}$ and the lowest $S_\text{OM}$ value was observed at 0.4 ps.
These are consistent with the LD results in which all 118 transitions that crossed barrier C occurred within 1.1 ps 
and their most probable transition time was 0.7 ps (the inset of Fig~\ref{fig:alanine_dipeptide}B).
However, when $t > 0.8$ ps, Path3, which passes the fully extended conformation region 
($\Phi$, $\Psi$) = (-180$^{\circ}$, 180$^{\circ}$) and barrier A and B becomes the pathway with the second lowest $S_\text{OM}$.
%%Among the transitions observed in LD simulations whose 
From the LD simulations, when $t>0.9$ ps, 25 pathways similar to Path3 were identified, which makes them the second dominant pathway.
These results demonstrate that the profile of $S_\text{OM}$ values is consistent with 
the distributions of transition times obtained from the LD simulations.
Note that the most probable transitions times observed from the LD simulations are longer than 
the minimum action transition times obtained from the CSA simulations.
This is because high-frequency motions due to thermal fluctuations are filtered out in the minimum action pathways~\cite{Olender1996,Elber1999,Elber2016}.
This means that the dwell time is well filtered out in the simulation, where a physically sufficient sampling time is assumed. 

The second example is finding possible pathways for the conformational change of hexane from the all-gauche(-) (g-g-g-) to the all-gauche(+) state (g+g+g+).
We assessed the sampling ability of Action-CSA by investigating how many pathways are found.
If we assume that dihedral angles do not cross a high barrier around the \emph{cis} state, 
%all possible transition pathways between the two all-gauche states can be enumerated (Table \ref{Sptab:hexane_pathway_list}).
all possible transition pathways between the two all-gauche states can be enumerated (Table \ref{tab:hexane_pathway_list}).
For this reaction, there exist 44 possible pathways in total.
If the symmetries of dihedral angles and atomic order are considered, all 44 pathways can be categorized into 14 unique pathway types.
We repeated the Action-CSA calculation of the reaction 40 times by using 200 trial pathways consisting of 100 replicas and a transition time of 3 ps.

\begin{figure}[!htbp]
  \includegraphics[width=0.70\columnwidth]{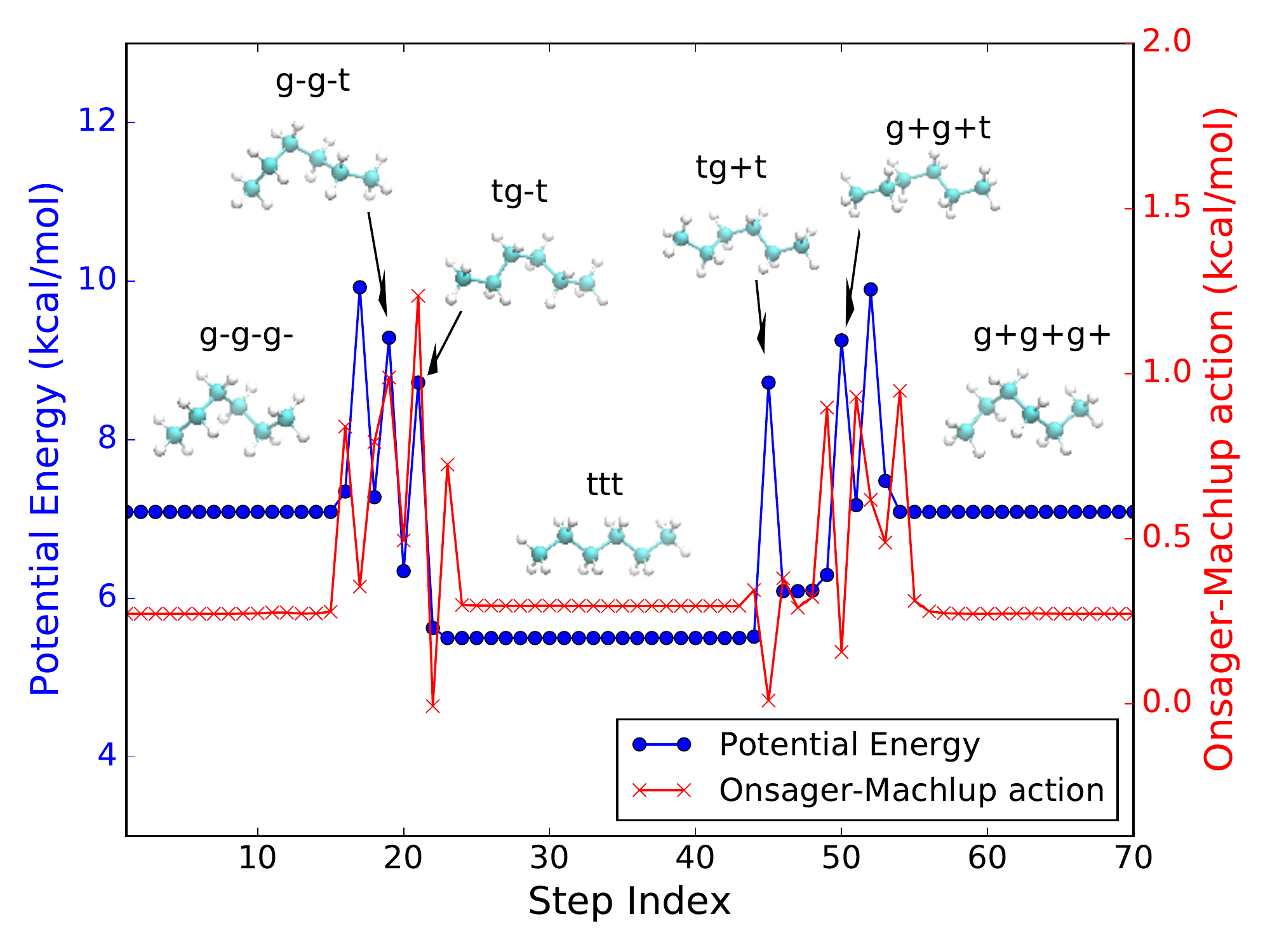} 
  \caption{The changes of potential energy and Onsager-Machlup action along the lowest action pathway 
    between the all-gauche(-) to the all-gauche(+) conformations of hexane in the vacuum, the C+C+ pathway.}
  \label{fig:hexane}
\end{figure}

In all simulations, the 6 lowest action pathways, C+C+, C-C-, T+C+, T+C-, C+M+, and C+M-, were found consistently.
The other higher action pathways except for the highest action pathway, M+XM-, were found in at least 29 out of 40 CSA simulations.
Only M+XM- was found in 9 simulations.
On average, a single CSA simulation sampled 12 out of 14 unique path types and 26 out of 44 possible pathways.
These results show that Action-CSA assuredly samples a number of lowest action, most dominant, pathways. 
The majority of the remaining pathways with higher actions can also be found with a tendency that relatively lower action pathways are more likely to be found. 
The sampling ability of Action-CSA can be improved by using a larger bank size.
The potential energy landscape of the C+C+ pathway corresponding to the least $S_\text{OM}$ shows that hexane crosses 6 energy barriers (Fig.~\ref{fig:hexane}).

%%%%%%%%%%%%%%%%%%%%%%%%%%%%%%%%%%%%%%%%%%%%%%%
%%%%%%%%%%%%%%%% FSD-1 example ? %%%%%%%%%%%%%%
%%%%%%%%%%%%%%%%%%%%%%%%%%%%%%%%%%%%%%%%%%%%%%%
The third example is finding the folding pathway of FSD-1, a 28-residue mini-protein 
that has been widely investigated as a model system for studying protein folding~\cite{Jang2002,Lei2004,Lei2006,Wu2010,Lei2009,Lee2012c}.%Sadqi2009}.
Folding pathways of FSD-1 from the fully extended conformation to the native structure 
were represented by using 100 replicas, a total folding time of 10 ps, and a temperature of 300 K.
The protein was represented by the PARAM19 force field~\cite{Neria1996} 
and solvation effects were considered using the FACTS implicit solvent model~\cite{Haberthur2008}.

\begin{figure}[!htbp]
  \includegraphics[width=0.70\columnwidth]{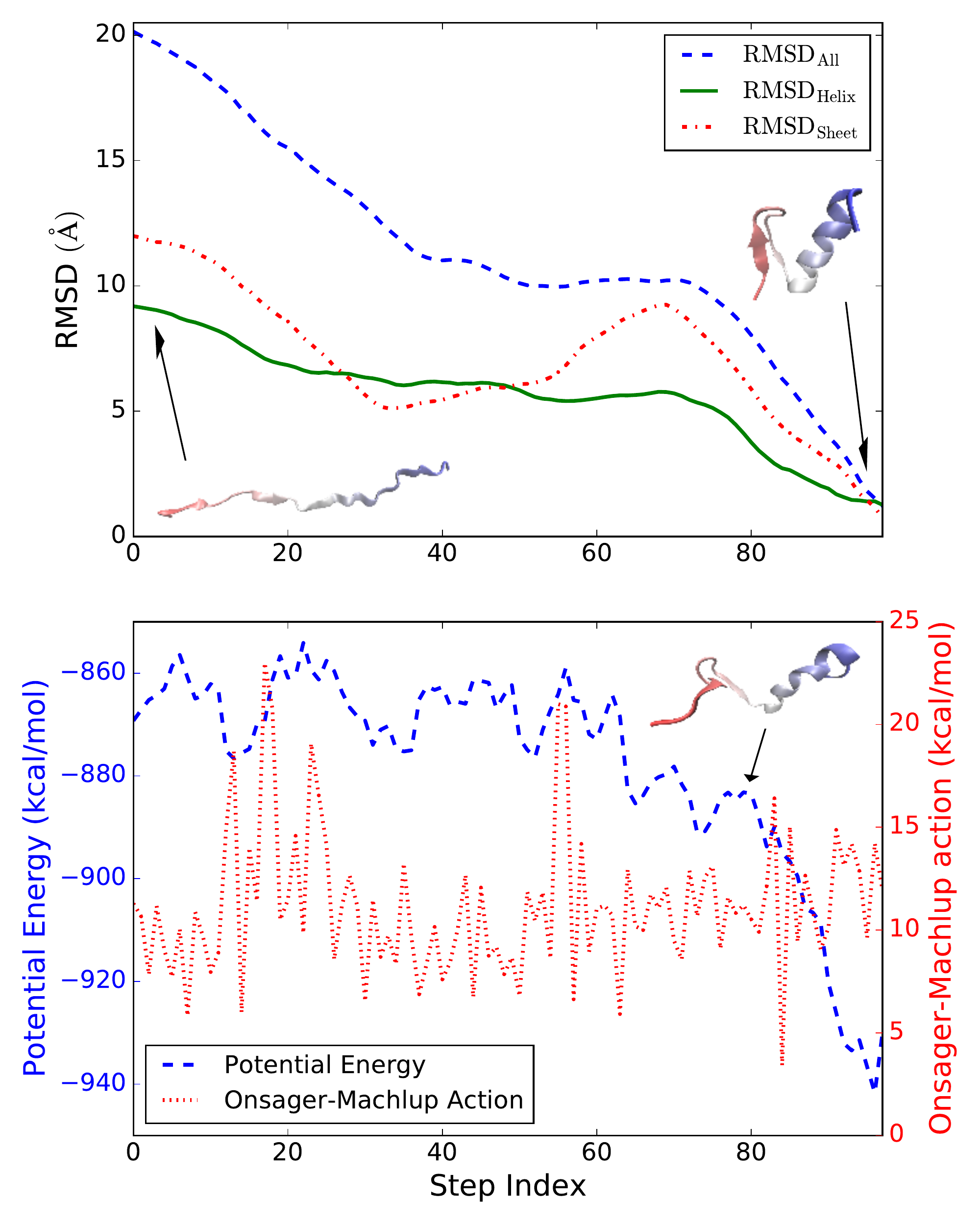}
  \caption{Upper panel: The RMSD values of the entire FSD-1 (blue), the C-terminal $\alpha$-helix (green, residue 14-28), 
    and the N-terminal $\beta$-hairpin (red, residue 1-13) from the native structure along the folding pathway. 
    Lower panel: The evolutions of potential energy (blue) and the Onsager-Machlup action (red) of FSD-1 along the folding pathway.}
  \label{fig:fsd1}
\end{figure}

The characteristics of the identified lowest action folding pathway are consistent with experiments where the N-terminal $\beta$-hairpin is more flexible than the C-terminal $\alpha$-helix~\cite{Feng2009a}.
A comparison of the RMSD values indicates that the $\alpha$-helix folds first.
Afterward, the folding of $\beta$-hairpin and the formation of hydrophobic core occur concurrently (the upper panel of Fig.~\ref{fig:fsd1}).
The potential energy landscape of FSD-1 folding shows that the potential energy decreases quickly after the \nth{80} step suggesting that the step may be the transition state of folding (the lower panel of Fig.~\ref{fig:fsd1}).
The conformation at the \nth{80} step shows that the helix is almost formed while the C-terminal region is not folded and the hydrophobic core is partially exposed, which is similar to transition states observed in previous conventional MD simulations~\cite{Jang2002,Lei2004,Lei2006,Wu2010}.

%%%%%%%%%%%%%%%%%%%%%%%%%%%%%%%%%%%%%%%%%%%%%%%
%%%%% Conclusion %%%%%%%%%%%%%%%%%%%%%%%%%%%%%%
%%%%%%%%%%%%%%%%%%%%%%%%%%%%%%%%%%%%%%%%%%%%%%%
In conclusion, we demonstrated that efficient global optimization of Onsager-Machlup action reveals multiple reaction pathways successfully. 
In this work, we introduced a new computational method that samples not only the most dominant pathway 
but also other possible pathways by optimizing Onsager-Machlup action using the CSA method.
The advantages of our method over existing pathway sampling methods are the fact 
that it samples multiple pathways regardless of the quality of initial guesses on pathways;
 it requires only the calculation of first derivatives; and 
 its sampling ability is not limited by high energy barriers separating pathways, 
%%that a) it samples multiple pathways without prior knowledge of pathways,
%that a) it does not depend on the quality of an initially guessed pathway, 
%2) it requires only the calculation of gradients, and 
%3) its sampling ability is not limited by high energy barriers between different pathways, 
which is a major limiting factor of previous MD-based pathway sampling methods in exploring pathway space~\cite{Eastman2001,Faccioli2006,Mathews2006,Sega2007,Fujisaki2010,Beccara2012,Fujisaki2013}.
In addition, we identified that the profile of minimum Onsager-Machlup actions found with different transition time parameters 
provide kinetic information on pathways.
%%provide information on the relationship between transition time and the dominance of pathways.
In terms of implementation, Action-CSA calculation is massively parallel because the local minimization of each trial pathway is independent of each other.
Thus, pathway samplings for larger systems are readily possible with help of a large cluster system.

\begin{center}
  \begin{longtable}{|p{3cm}|p{12cm}|}
  \caption{List of 14 unique pathway types and 44 non redundant pathways for conformational change of hexane from g-g-g- to g+g+g+.} 
  \label{tab:hexane_pathway_list}
  \endfirsthead
  \endhead
  %\begin{tabular}[t]{|c|c|}
  \hline
  Unique path type & Non redundant path \\ 
  \hline
 C+C+ &	g-g-g-  $\rightarrow$  tg-g-  $\rightarrow$  tg-t  $\rightarrow$  ttt  $\rightarrow$  tg+t  $\rightarrow$  tg+g+  $\rightarrow$  g+g+g+ \\
      & g-g-g-  $\rightarrow$  g-g-t  $\rightarrow$  tg-t  $\rightarrow$  ttt  $\rightarrow$  tg+t  $\rightarrow$  g+g+t  $\rightarrow$  g+g+g+ \\ \hline
 C+C- &	g-g-g-  $\rightarrow$  g-g-t  $\rightarrow$  tg-t  $\rightarrow$  ttt  $\rightarrow$  tg+t  $\rightarrow$  tg+g+  $\rightarrow$  g+g+g+ \\
      & g-g-g-  $\rightarrow$  tg-g-  $\rightarrow$  tg-t  $\rightarrow$  ttt  $\rightarrow$  tg+t  $\rightarrow$  g+g+t  $\rightarrow$  g+g+g+ \\ \hline
 T+C+ &	g-g-g-  $\rightarrow$  tg-g-  $\rightarrow$  tg-t  $\rightarrow$  ttt  $\rightarrow$  ttg+  $\rightarrow$  tg+g+  $\rightarrow$  g+g+g+ \\
      & g-g-g-  $\rightarrow$  g-g-t  $\rightarrow$  tg-t  $\rightarrow$  ttt  $\rightarrow$  g+tt  $\rightarrow$  g+g+t  $\rightarrow$  g+g+g+ \\
      & g-g-g-  $\rightarrow$  g-g-t  $\rightarrow$  g-tt  $\rightarrow$  ttt  $\rightarrow$  tg+t  $\rightarrow$  g+g+t  $\rightarrow$  g+g+g+ \\
      & g-g-g-  $\rightarrow$  tg-g-  $\rightarrow$  ttg-  $\rightarrow$  ttt  $\rightarrow$  tg+t  $\rightarrow$  tg+g+  $\rightarrow$  g+g+g+ \\ \hline
 T+C- & g-g-g-  $\rightarrow$  g-g-t  $\rightarrow$  tg-t  $\rightarrow$  ttt  $\rightarrow$  ttg+  $\rightarrow$  tg+g+  $\rightarrow$  g+g+g+ \\
      & g-g-g-  $\rightarrow$  tg-g-  $\rightarrow$  tg-t  $\rightarrow$  ttt  $\rightarrow$  g+tt  $\rightarrow$  g+g+t  $\rightarrow$  g+g+g+ \\
      & g-g-g-  $\rightarrow$  g-g-t  $\rightarrow$  g-tt  $\rightarrow$  ttt  $\rightarrow$  tg+t  $\rightarrow$  tg+g+  $\rightarrow$  g+g+g+ \\
      & g-g-g-  $\rightarrow$  tg-g-  $\rightarrow$  ttg-  $\rightarrow$  ttt  $\rightarrow$  tg+t  $\rightarrow$  g+g+t  $\rightarrow$  g+g+g+ \\ \hline
 C+M+ & g-g-g-  $\rightarrow$  tg-g-  $\rightarrow$  tg-t  $\rightarrow$  ttt  $\rightarrow$  ttg+  $\rightarrow$  g+tg+  $\rightarrow$  g+g+g+ \\
      & g-g-g-  $\rightarrow$  g-g-t  $\rightarrow$  tg-t  $\rightarrow$  ttt  $\rightarrow$  g+tt  $\rightarrow$  g+tg+  $\rightarrow$  g+g+g+ \\
      & g-g-g-  $\rightarrow$  g-tg-  $\rightarrow$  ttg-  $\rightarrow$  ttt  $\rightarrow$  tg+t  $\rightarrow$  g+g+t  $\rightarrow$  g+g+g+ \\ 
      & g-g-g-  $\rightarrow$  g-tg-  $\rightarrow$  g-tt  $\rightarrow$  ttt  $\rightarrow$  tg+t  $\rightarrow$  tg+g+  $\rightarrow$  g+g+g+ \\ \hline
 C+M- & g-g-g-  $\rightarrow$  g-g-t  $\rightarrow$  tg-t  $\rightarrow$  ttt  $\rightarrow$  ttg+  $\rightarrow$  g+tg+  $\rightarrow$  g+g+g+ \\
      & g-g-g-  $\rightarrow$  tg-g-  $\rightarrow$  tg-t  $\rightarrow$  ttt  $\rightarrow$  g+tt  $\rightarrow$  g+tg+  $\rightarrow$  g+g+g+ \\
      & g-g-g-  $\rightarrow$  g-tg-  $\rightarrow$  g-tt  $\rightarrow$  ttt  $\rightarrow$  tg+t  $\rightarrow$  g+g+t  $\rightarrow$  g+g+g+ \\
      & g-g-g-  $\rightarrow$  g-tg-  $\rightarrow$  ttg-  $\rightarrow$  ttt  $\rightarrow$  tg+t  $\rightarrow$  tg+g+  $\rightarrow$  g+g+g+ \\ \hline
 T+T+ & g-g-g-  $\rightarrow$  tg-g-  $\rightarrow$  ttg-  $\rightarrow$  ttt  $\rightarrow$  ttg+  $\rightarrow$  tg+g+  $\rightarrow$  g+g+g+ \\ 
      & g-g-g-  $\rightarrow$  g-g-t  $\rightarrow$  g-tt  $\rightarrow$  ttt  $\rightarrow$  g+tt  $\rightarrow$  g+g+t  $\rightarrow$  g+g+g+ \\ \hline
 T+T- & g-g-g-  $\rightarrow$  tg-g-  $\rightarrow$  ttg-  $\rightarrow$  ttt  $\rightarrow$  g+tt  $\rightarrow$  g+g+t  $\rightarrow$  g+g+g+ \\ 
      & g-g-g-  $\rightarrow$  g-g-t  $\rightarrow$  g-tt  $\rightarrow$  ttt  $\rightarrow$  ttg+  $\rightarrow$  tg+g+  $\rightarrow$  g+g+g+ \\ \hline
 T+M+ & g-g-g-  $\rightarrow$  g-g-t  $\rightarrow$  g-tt  $\rightarrow$  ttt  $\rightarrow$  g+tt  $\rightarrow$  g+tg+  $\rightarrow$  g+g+g+ \\
      & g-g-g-  $\rightarrow$  tg-g-  $\rightarrow$  ttg-  $\rightarrow$  ttt  $\rightarrow$  ttg+  $\rightarrow$  g+tg+  $\rightarrow$  g+g+g+ \\
      & g-g-g-  $\rightarrow$  g-tg-  $\rightarrow$  g-tt  $\rightarrow$  ttt  $\rightarrow$  g+tt  $\rightarrow$  g+g+t  $\rightarrow$  g+g+g+ \\ 
      & g-g-g-  $\rightarrow$  g-tg-  $\rightarrow$  ttg-  $\rightarrow$  ttt  $\rightarrow$  ttg+  $\rightarrow$  tg+g+  $\rightarrow$  g+g+g+ \\ \hline
 T+M- & g-g-g-  $\rightarrow$  g-g-t  $\rightarrow$  g-tt  $\rightarrow$  ttt  $\rightarrow$  ttg+  $\rightarrow$  g+tg+  $\rightarrow$  g+g+g+ \\
      & g-g-g-  $\rightarrow$  tg-g-  $\rightarrow$  ttg-  $\rightarrow$  ttt  $\rightarrow$  g+tt  $\rightarrow$  g+tg+  $\rightarrow$  g+g+g+ \\
      & g-g-g-  $\rightarrow$  g-tg-  $\rightarrow$  g-tt  $\rightarrow$  ttt  $\rightarrow$  ttg+  $\rightarrow$  tg+g+  $\rightarrow$  g+g+g+ \\
      & g-g-g-  $\rightarrow$  g-tg-  $\rightarrow$  ttg-  $\rightarrow$  ttt  $\rightarrow$  g+tt  $\rightarrow$  g+g+t  $\rightarrow$  g+g+g+ \\ \hline
 M+M- & g-g-g-  $\rightarrow$  g-tg-  $\rightarrow$  g-tt  $\rightarrow$  ttt  $\rightarrow$  ttg+  $\rightarrow$  g+tg+  $\rightarrow$  g+g+g+ \\
      & g-g-g-  $\rightarrow$  g-tg-  $\rightarrow$  ttg-  $\rightarrow$  ttt  $\rightarrow$  g+tt  $\rightarrow$  g+tg+  $\rightarrow$  g+g+g+ \\
      & g-g-g-  $\rightarrow$  g-tg-  $\rightarrow$  g-tt  $\rightarrow$  ttt  $\rightarrow$  g+tt  $\rightarrow$  g+tg+  $\rightarrow$  g+g+g+ \\
      & g-g-g-  $\rightarrow$  g-tg-  $\rightarrow$  ttg-  $\rightarrow$  ttt  $\rightarrow$  ttg+  $\rightarrow$  g+tg+  $\rightarrow$  g+g+g+ \\ \hline
T+XT- & g-g-g-  $\rightarrow$  g-g-t  $\rightarrow$  g-tt  $\rightarrow$  g-tg+ $\rightarrow$  ttg+  $\rightarrow$  tg+g+  $\rightarrow$  g+g+g+ \\ 
      & g-g-g-  $\rightarrow$  tg-g-  $\rightarrow$  ttg-  $\rightarrow$  g+tg- $\rightarrow$  g+tt  $\rightarrow$  g+g+t  $\rightarrow$  g+g+g+ \\ \hline
 TXM  & g-g-g-  $\rightarrow$  g-tg-  $\rightarrow$  g-tt  $\rightarrow$  g-tg+ $\rightarrow$  ttg+  $\rightarrow$  tg+g+  $\rightarrow$  g+g+g+ \\
      & g-g-g-  $\rightarrow$  tg-g-  $\rightarrow$  ttg-  $\rightarrow$  g+tg-  $\rightarrow$  g+tt  $\rightarrow$  g+tg+  $\rightarrow$  g+g+g+ \\
      & g-g-g-  $\rightarrow$  g-g-t  $\rightarrow$  g-tt  $\rightarrow$  g-tg+ $\rightarrow$  ttg+  $\rightarrow$  g+tg+  $\rightarrow$  g+g+g+ \\
      & g-g-g-  $\rightarrow$  g-tg-  $\rightarrow$  ttg-  $\rightarrow$  g+tg-  $\rightarrow$  g+tt  $\rightarrow$  g+g+t  $\rightarrow$  g+g+g+ \\ \hline
M+XM- & g-g-g-  $\rightarrow$  g-tg-  $\rightarrow$  g-tt  $\rightarrow$  g-tg+ $\rightarrow$  ttg+  $\rightarrow$  g+tg+  $\rightarrow$  g+g+g+ \\
      & g-g-g-  $\rightarrow$  g-tg-  $\rightarrow$  ttg-  $\rightarrow$  g+tg-  $\rightarrow$  g+tt  $\rightarrow$  g+tg+  $\rightarrow$  g+g+g+ \\ \hline
\hline
%\end{tabular}
\end{longtable}
\end{center}

\begin{acknowledgments}
  The authors wish to acknowledge helpful discussions with Attila Szabo.
  The research was supported by the Intramural Research Program of the NIH, NHLBI. 
  Computational resources and services used in this work were provided by the LoBoS cluster of the National Institutes of Health.
\end{acknowledgments}

\bibliography{all}

\end{document}